# The problematic nature of potentially polynomial-time algorithms solving the subset-sum problem

Antonios Syreloglou[1]

**Abstract** The main purpose of this paper is to study the NP-complete subset-sum problem, not in the usual context of time-complexity-based classification of the algorithms (exponential/polynomial), but through a new kind of algorithmic classification which we introduce based on a property that all known exponential-time algorithms share. We then construct a theoretical mathematical environment within which we compare the two classes that are produced from the new classification; one class is characterized by a normal mathematical nature, whereas the other by a problematic one. These results are transferred to exponential/polynomial algorithms, through a conjecture that links the two classifications. As for the mathematical environment, it consists of a simple random experiment designed in such a way that the algorithmic operation is linked to it. We study the random experiment with a stochastic process which is the main tool for the comparison of the two classes.

## 1 Introduction

First of all, let us give a short description of the subset-sum problem. Given a set $S = \{x_1, x_2, \ldots, x_n\}$ of $n$ integers, the objective is to find if there is a non-empty subset that sums to zero; a subset that sums to zero will be called subset-solution.

The subset-sum problem belongs to the class of NP-complete problems which hold a key role in computational complexity theory. Regarding the solution of the problem, it is known that only exponential-time algorithms exist so far. Whether a polynomial-time algorithm exists is the essence of one of the most famous problems in theoretical computer science, the P versus NP problem. For more information about NP-complete problems, complexity classes P and NP, the time complexity of algorithms (exponential/polynomial) and the P versus NP problem, the reader may see [1].

This paper revolves around the meaning and the notion of the phrase "an algorithm determines whether a subset is a subset-solution or not" or simply "an algorithm determines a subset". An algorithm may determine this through any necessary operation it is designed to perform; from "very tangible" operations, like computing the sum of a certain subset, to "intangible" ones, like just reading the $n$ integers and "magically" outputting the answer "no, there is not a subset-solution", thus having determined that none of the subsets is a subset-solution (such a "magical" operation is obviously not known to exist). In the latter case, note that by outputting the answer "yes, there is a subset-solution" without being able to name this subset-solution, the algorithm would not have determined anything yet the problem would have been solved.

In this paper, we study the algorithms which solve the subset-sum problem by determining all subsets of $S = \{x_1, x_2, \ldots, x_n\}$, meaning that each subset of $S$ is determined as to whether it is a subset-solution. The set $S$, which we input into such an algorithm to determine its subsets, will be accordingly called input-set. Furthermore, the study of algorithms is not done through the usual definitions using big-$O$ notation, but through a property that is derived from the way subset-sum algorithms work. In particular, this property – which is described in the next paragraphs and we will be calling it property $P_1$ – is found in all known exponential-time algorithms. Then, we present an appropriately modified version of property $P_1$, called property $P_2$. But since no algorithm is known to satisfy property $P_2$, we introduce an imaginary class of algorithms equipped with it. So, we have a new kind of classification: the class of algorithms that satisfy property $P_1$ ($P_1$-algorithms) and the – imaginary – class of algorithms that satisfy property $P_2$ ($P_2$-algorithms) (in Chapter 2).

---
1 Contact information: a.syrelog@gmail.com

Let us note that through this new classification we do not make a direct comparison between the general class of exponential-time algorithms and that of polynomial-time ones. However, we do attempt to "shed some mathematical light" to such a comparison by doing the following. We show that $P_2$-algorithms are characterized by a problematic mathematical nature, whereas $P_1$-algorithms (including all known exponential-time algorithms) are not (in Chapter 4). But we have conjectured, based on certain reasoning (in Chapter 3), that the class of $P_2$-algorithms includes the class of polynomial-time algorithms which determine all subsets of the input-set and so, polynomial-time algorithms are characterized by this problematic nature as well. We believe that this nature of $P_2$-algorithms (and potentially polynomial-time algorithms, according to the conjecture) exactly agrees with the fact that no such algorithm is known to exist.

As far as the demonstration of this algorithmic nature is concerned, the way to do it is by creating a theoretical mathematical environment within which we compare the two classes, the $P_1$ and $P_2$-algorithms. This mathematical environment consists of a simple random experiment designed in such a way that the operation of the algorithm is linked to it. Then, we study how the random experiment evolves over an axis representing the probability of an appropriately chosen event. To do this, we introduce a stochastic process with an index set created in such a way to serve this unusual[2] kind of axis. This stochastic process is the main tool for both the normal and the problematic behavior (of $P_1$ and $P_2$-algorithms respectively) to emerge. For more information about the theory of stochastic processes, the reader may see [2].

## 2 The new algorithmic classification

In this section, we describe the new way of classifying subset-sum algorithms. As stated in the introduction, this classification is achieved by observing a property found in the algorithmic operation of all known exponential-time algorithms. In order to talk about this property, we need the following definition.

**Definition 2.1** Let $S$ be an input-set. We say that an algorithm checks a subset $\sigma$ of $S$ when the algorithm performs the final necessary operation to determine $\sigma$ independently of other subsets of $S$. A step in which checking takes place is called main step.

The definition is as complete as needed for the theory developed in this paper. The term "step" is used as it is perceived in a general definition of an algorithm, i.e. "an algorithm is a step-by-step set of operations to be performed" or "an algorithm is a finite, precise list of precise steps". Also, when we say "operation" we do not necessarily mean an arithmetic one, but every kind of algorithmic operation. Now, let us present an example to see exactly how the definition is applied.

*Example 2.2* The brute-force exponential-time algorithm.
Here, the algorithm cycles through all possible subsets of the $n$ integers and, for every one of them, determines whether the subset is a subset-solution or not. A less rigorous way of presenting the algorithm would suggest that the running time is of order $O(2^n)$ ; in each step a different subset is checked, since the algorithm determines each subset by adding its elements together and, therefore, independently of other subsets. In that case, all steps would be main steps.

---

[2] With the usual ones historically being time (mainly) or space.

In a rigorous context, there are $2^n - 1$ subsets and, to examine each subset, the algorithm needs to sum at most $n$ elements. Then, the running time is of order $O(2^n n)$. Taking under consideration definition 2.1, the main steps are the steps where the last element of each subset is added to the sum of all the other elements. Indeed, adding the last element constitutes the final necessary operation to determine independently of other subsets. But let us clarify exactly why this operation is compatible with independence. Typically, the last element itself is a subset of $S = \{x_1, x_2, \ldots, x_n\}$. And since it is added to the sum of all the other elements of the original subset, one could say that the result depends on it, meaning on another subset; but this is false. What is going on is that the result does not depend on the actual (value of the) last element, but on how the last element is related to the other elements. Indeed, in order for the subset to be determined, the information obtained from the addition of the last element is useless on its own unless the sum of all the other elements is known. Therefore, the contribution of the last element is not that of an individual, separate subset but that of an element of the original subset. The same is true for each element added to the sum before the last element. As a result, the algorithm determines the original subset independently of other subsets.

For an application of the above, let us consider subset $\{x_1, x_4, x_7\}$ of $S = \{x_1, x_2, \ldots, x_n\}$. If the brute-force algorithm calculates the sum $x_1 + x_4 = a$ in the first step and the sum $a + x_7$ in the second step, the subset is checked in the second step (main step) as it is only then that the algorithm can determine whether the subset is a subset-solution, having determined it independently of other subsets. The first step could be considered as a procedural step.

As more examples follow, we will see that the most important feature of the definition is the phrase "determine independently of other subsets". We say "the final necessary operation" performed by the algorithm so that we can pinpoint the step in which a subset is determined, but it is the entire procedure of this subset being determined that has to be done independently of other subsets, so that we can say the subset is checked (see example 2.2 where we add all the elements together).

We can, now, formally state the property and then, discuss it in details.

*Property $P_1$* : *Only one subset is checked in a main step.*

We call the class of algorithms that satisfy the above property $P_1$-algorithms. As we have already mentioned, property $P_1$ is satisfied by all known exponential-time algorithms, therefore they are $P_1$-algorithms (more details in the following examples). In a main step, where checking takes place according to definition 2.1, a $P_1$-algorithm checks a subset. The way such an algorithm is designed allows no more than this one subset to be checked. In the non-main steps, no subset is checked (by definition).

So, the fundamental purpose of a $P_1$-algorithm, which is to determine all subsets of the input-set, is achieved by checking at most one subset in a single step. Now, let us present certain examples of $P_1$-algorithms.

*Example 2.2* The brute-force exponential-time algorithm (continuation).
By the way this algorithm works, it is obvious that only one subset is checked in a single, main step.

*Example 2.3* The "Horowitz and Sahni" exponential-time algorithm.
For detailed information on how this algorithm works, the reader may see [3]. The algorithm splits arbitrarily the $n$ integers (of the input-set) into two sets of $n/2$ each and for each set, it stores the sums of all possible subsets of the $n/2$ integers of the set in a list, which is then sorted. So, the algorithm generates two sorted lists with $2^{n/2} - 1$ elements each. Given the two lists, the algorithm passes through the first list in decreasing order (starting at the largest element) and the second list in increasing order (starting at the smallest element). Whenever the sum of the current couple of elements (one element from the first list and one from the second list, combined for a new subset) is more than zero, the algorithm moves to the next element in the first list. If it is less than zero, it moves to the next element in the second list.

During the process of calculating the sums (of all possible subsets of the $n/2$ integers) to be stored in each list, property $P_1$ is obviously satisfied exactly like in example 2.2. Now, let us see why the property is satisfied by the rest of the algorithm as well. If the sum of the current couple of elements is found to be more than zero, the algorithm will move to the next element in the first list. There is no need of examining the combinations of the current element from the first list with the larger elements in the second list, since the resulting subsets must sum to even larger numbers and, therefore, none of them can be a subset-solution. In that way, it is not only the current couple of elements that is determined in the step where its sum is calculated, but also (in the same step) all the above subsets-combinations. This does not contradict property $P_1$, since only the current couple is checked. The subsets-combinations are determined but not checked. Indeed, according to definition 2.1 the current couple is the only subset that the algorithm determines in a way completely independent of other subsets, by calculating its sum that is. On the other hand, the subsets-combinations (each one of them) are determined in a way that is clearly dependent on the current couple; the algorithm determines that none of these subsets-combinations is a subset-solution based on the fact that this specific current couple sums to a number larger than zero. Therefore, many subsets are determined in the step but only one of them (the current couple of elements) is checked. Respectively for the case in which the sum of the current couple is less than zero, there is no need of examining the combinations of the current element from the second list with the smaller elements in the first list, since the resulting subsets must sum to even smaller numbers.

Through example 2.3, it becomes clear that the subsets-combinations play an important role in the process of determining all possible subsets. The term "subsets-combinations" itself is informal, just for the needs of this specific example. But in order to refer to such kind of subsets in all algorithms, we need a formal definition.

**Definition 2.4** Let us have an algorithm and let $\sigma$ be a subset of an input-set $S$. We call $\sigma$ a collateral gain if it is determined by the algorithm but not checked.

Since checking is performed independently of other subsets, an obviously equivalent definition to 2.4 is the following.

**Definition 2.5** Let us have an algorithm and let $\sigma$ be a subset of an input-set $S$. We call $\sigma$ a collateral gain if it is determined by the algorithm in a way which is dependent on other (one or more) subsets. Let these other subsets be $\sigma_1, \sigma_2, \ldots, \sigma_k$, then "other" is defined as $\sigma \not\equiv \sigma_1, \sigma_2, \ldots, \sigma_k$.

In the above definition, we defined "other" using the notation $\not\equiv$ instead of $\neq$. This is important for the following reason. When two subsets $a$ and $b$ are not equal ($\neq$), then obviously they are not identical ($\not\equiv$) either. In other words, "$a \neq b \Rightarrow a \not\equiv b$". But in the context of an algorithm, $a$ and $b$ may be equal (they may coincide) but not identical, as a result of the algorithm viewing these two subsets as being different (let us wait for following example 2.7 to see this in application). In other words, "$a = b \not\Rightarrow a \equiv b$". Therefore, if we used $\neq$ in the definition and had $a = b$, subset $b$ would not be classified as an "other" subset to $a$, in contrast to what the algorithmic context could suggest ($a \not\equiv b$). To summarize, we are interested in defining "other" semantically (expressed by $\not\equiv$) and not syntactically (expressed by $\neq$).

So, the subsets-combinations determined in the "Horowitz and Sahni" algorithm are collateral gains. They are determined in a way that is dependent on the current couple, which is clearly an "other" subset. Indeed, none of the subsets-combinations is equal to the current couple, thus none is identical to it.

Let us explain the inspiration behind the selection of the term "collateral gain". A subset called collateral gain is obviously a gain for an algorithm, as it is a subset being determined, and it is collateral because the presence of other subsets is necessary and crucial for the subset to be determined (given that it is determined in a way which is dependent on these other subsets).

Before we present the next example, we make the following observation.

**Corollary 2.6** *A subset which is determined by an algorithm is either a checked subset or a collateral gain.*

This corollary is derived from the law of excluded middle (or principle of the excluded third).

*Example 2.7* The dynamic programming (pseudo-polynomial time) algorithm.

The algorithm makes use of a boolean-valued (True or False) function $F(i, s)$ as the answer to "there is a subset of $x_1, \ldots, x_i$ that sums to $s$", with the value of $F(n, 0)$ being the solution of the decision version of the problem. Let us give a short description of the way the algorithm works. Let $A$ be the sum of the positive elements and $B$ the sum of the negative elements of the input-set. An array is created, so that all values $F(i, s)$ can be stored for $1 \leq i \leq n$ and $B \leq s \leq A$. Initially, for $i = 1$ and $B \leq s \leq A$, the algorithm sets the value of $F(i, s)$ to be equal to the boolean value of the expression $(x_i = s)$. Then, for $i = 2, 3, \ldots, n$ and $B \leq s \leq A$, the algorithm computes recursively the values $F(i, s)$ according to

$$F(i, s) = F(i - 1, s) \lor (x_i = s) \lor F(i - 1, s - x_i).$$

Also, $F(i, s) = $ False if $s < B$ or $s > A$. The time of the algorithm is $O(n(A - B))$, since we have $1 \leq i \leq n$ and $B \leq s \leq A$, resulting in $n(A - B)$ values of $F(i, s)$. Therefore, let us consider that the computation of a single value of $F(i, s)$ is done in a single step. Finally, a modification to the algorithm, so that the subset which is a subset-solution can be named at any point, can easily be made without changing anything in the general operation or running time of the algorithm. Because of this ability to name the subset-solution (in contrast to the example we saw in the introduction), the following two phrases are equivalent in meaning: "the algorithm answers that there is/is not a subset of $x_1, \ldots, x_i$ that sums to 0 (subset-solution)" and "the algorithm determines the subsets of $x_1, \ldots, x_i$ ".

Now, let us describe what happens in $F(i, 0)$-steps, the steps where $F(i, 0)$ is computed by the algorithm for $i = 1, 2, 3, \ldots, n$. For $i = 1$, the algorithm sets $F(1,0) = (x_1 = 0)$ and by doing so, determines independently of other subsets (as no subset "participates" in this operation other than $\{x_1\}$) whether $\{x_1\}$ is a subset-solution. Therefore, a checking takes place, which makes the step (the $F(1,0)$-step) a main step. And the subset $\{x_1\}$ is the only subset that the algorithm checks in this step.

For $i = 2, 3, \ldots, n$, the algorithm computes the expression

$$F(i, 0) = F(i - 1, 0) \lor (x_i = 0) \lor F(i - 1, -x_i),$$

each $i$ in a different step. Let us analyze this expression:

(1) If $F(i - 1, 0) = $ True, then $F(i, 0) = $ True. The algorithm answers that there is a subset of $x_1, \ldots, x_i$ – among those that do not contain $x_i$ as their element – which sums to 0 (subset-solution) since there is a subset of $x_1, \ldots, x_{i-1}$ that sums to 0. In other words, the algorithm determines the subsets of $x_1, \ldots, x_i$ that do not contain $x_i$ as collateral gains, in a way that is dependent on the subsets of $x_1, \ldots, x_{i-1}$. (Although they coincide, note that the subsets of $x_1, \ldots, x_i$ that do not contain $x_i$ belong to the set $C_i$ of all subsets of $x_1, \ldots, x_i$ ,

[5]

whereas the subsets of $x_1, \ldots, x_{i-1}$ form a different set $C_{i-1}$. In the context of this algorithm, the information that the algorithm needs – in order to determine – is transferred from $F(i-1, 0)$ to $F(i, 0)$, that is from set $C_{i-1}$ to set $C_i$. Since $C_{i-1} \neq C_i$, the subsets of $x_1, \ldots, x_{i-1}$ are not identical to the subsets of $x_1, \ldots, x_i$ that do not contain $x_i$; therefore, they are classified as other subsets.)

(2) If $F(i-1, -x_i) =$ True, then $F(i, 0) =$ True. Here, the algorithm answers that there is a subset of $x_1, \ldots, x_i$ – among those that contain $x_i$ (except $\{x_i\}$) – which sums to $0$ (subset-solution) since there is a subset of $x_1, \ldots, x_{i-1}$ that sums to $-x_i$ (and by adding $x_i$ to this subset it becomes a subset-solution) . Again in other words, the algorithm determines the subsets of $x_1, \ldots, x_i$ that contain $x_i$ (except $\{x_i\}$) as collateral gains, in a way that is dependent on the subsets of $x_1, \ldots, x_{i-1}$. (Note that this way is not dependent on $\{x_i\}$, although it may appear so in the form of $-x_i$. The reason is that after the information is transferred from set $C_{i-1}$ to set $C_i$ through $F(i-1, -x_i)$, the subsets that are determined contain $x_i$ and so, we have a situation similar to the "last element" in example 2.2 . As mentioned there, the contribution of $x_i$ is not that of an individual, separate subset but that of an element of each of the subsets that are determined.)

(3) If $F(i-1, 0) =$ False, then the algorithm answers that there is not a subset of $x_1, \ldots, x_i$ – among those that do not contain $x_i$ as their element – which sums to $0$ (subset-solution) since there is not a subset of $x_1, \ldots, x_{i-1}$ that sums to $0$. The rest is exactly the same as in (1).

(4) If $F(i-1, -x_i) =$ False, then the algorithm answers that there is not a subset of $x_1, \ldots, x_i$ – among those that contain $x_i$ (except $\{x_i\}$) – which sums to $0$ (subset-solution) since there is not a subset of $x_1, \ldots, x_{i-1}$ that sums to $-x_i$. The rest is exactly the same as in (2).

Therefore, through $F(i-1, 0)$ and $F(i-1, -x_i)$ combined, all subsets of $x_1, \ldots, x_i$ except $\{x_i\}$ are determined as collateral gains in a way that is dependent on the subsets of $x_1, \ldots, x_{i-1}$.

(5) Through $(x_i = 0)$ (True or False), which is the only operation left in the expression, the subset $\{x_i\}$ is determined in a way that is obviously independent of other subsets. Thus, the algorithm checks $\{x_i\}$ and $F(i, 0)$-step is a main step.

To summarize, each $F(i, 0)$-step for $i = 1, 2, 3, \ldots, n$ is a main step in which only one subset ($\{x_i\}$) is checked. As for all the other $F(i, s)$-steps ($s \neq 0$), the algorithm technically does not determine whether a subset is a subset-solution, since the values of $F(i, s)$ that are computed concern subsets that sum to $s \neq 0$. Even so, the operations in the $F(i, s)$-steps ($s \neq 0$) are exactly the same as in $F(i, 0)$-steps, therefore the analysis would be exactly the same as in the previous paragraphs. So, when we combine all conclusions together, we see that the algorithm is a $P_1$-algorithm.

Through the above examples, we described how known exponential-time algorithms operate and we saw that their operation ensures that the $P_1$-property is satisfied. The same can be said for all known exponential-time algorithms.

Now, it is clear that the property itself contains a restriction ("only one subset…") that appears to be linked with exponentiality. By lifting the restriction, we modify the property (as mentioned in the introduction) and this allows us to state a conjecture which agrees with our mathematical intuition. But firstly, let us present the new, modified property.

*Property $P_2$* : *There is at least one main step in which more than one subsets are checked.*

For example, an algorithm that checks two subsets in the same step satisfies property $P_2$. Let these two subsets be subset $a$ and subset $b$. For both subsets $a$ and $b$ to be checked in the same step, the algorithm must determine subset $a$ independently of other subsets including subset $b$ (so that $a$ is not a collateral gain of $b$) and also, must determine $b$ independently of other subsets including $a$.

No algorithm that satisfies property $P_2$ is known to exist. Therefore, the class of algorithms which satisfy property $P_2$ can only be an imaginary one. We call this class $P_2$-algorithms and, later in this paper, we compare it to $P_1$-algorithms.

Finally, let us present the following conjecture in our attempt to connect the imaginary class of $P_2$-algorithms with the polynomial-time algorithms. Besides the intuitive approach to formulating such a conjecture, the entire mathematical reasoning behind it is given in the next chapter.

**Conjecture 2.8** *The class of $P_2$-algorithms includes the class of polynomial-time algorithms which determine all subsets of an input-set.*

Conjecture 2.8 says that a polynomial-time algorithm which determines all subsets of an input-set is a $P_2$-algorithm. An illustration of 2.8 can be seen in following Fig. 1.

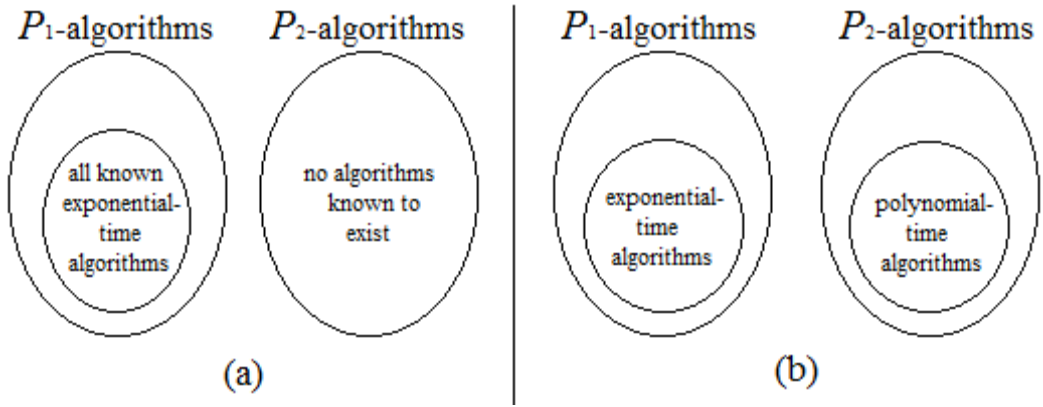

**Fig. 1** (a) The current state, (b) by assuming 2.8, with the additional (conjectured) information that the class of exponential-time algorithms is included in the class of $P_1$-algorithms.

## 3 The reasoning behind the conjecture

In this chapter, we develop the mathematical reasoning behind conjecture 2.8. But firstly, we need to state another conjecture that serves as an axiom on which we build and develop this entire reasoning. This new conjecture concerns collateral gains and it is fundamental, as we believe, to the way algorithms operate. Let us remind that a collateral gain is determined in a way which depends on other subsets. To simplify things, instead of saying that the way in which a collateral gain is determined depends on other subsets, we will be saying that the collateral gain depends on other subsets (in order to be determined).

**Conjecture 3.1** *A collateral gain depends only on determined subsets.*

Of course, conjecture 3.1 is true for all known exponential-time algorithms. We can easily see that, in example 2.3, the collateral gains (which are the subsets-combinations) depend on "the current couple", which is a determined subset. Also, in example 2.7, the collateral gains (which are all the subsets of $x_1, \dots, x_i$ except $\{x_i\}$) depend on the subsets of $x_1, \dots, x_{i-1}$, which are determined subsets (since the $F(i-1, 0)$ has already been computed, being a necessary step).

We believe that it is only from determined subsets (and not from undetermined ones) that meaningful information can be derived in order to produce collateral gains. An undetermined subset appears to be a subset that has not been manipulated by the algorithm at all, thus no information can be derived from it in order to be transferred to a potential collateral gain. Note that we refer to information in general. By that, we want to highlight the fact that whatever way the algorithm uses to determine a subset, it is always about obtaining information (through algorithmic operations) to achieve this.

[7]

*Remark 3.2* Definition 2.5 requires that a collateral gain must depend on at least one other subset. So, the definition does not exclude the case where a collateral gain depends on other subsets and on itself as well. But the fact that such a case is not known to exist in an algorithm agrees with the assumption of conjecture 3.1, since this case is not compatible with 3.1. Indeed, let us assume that a collateral gain depends on itself in order to be determined. This means that it depends on an undetermined subset, since it has not been determined yet. But depending on an undetermined subset is a contradiction if we have already assumed 3.1. Therefore, when assuming conjecture 3.1, a collateral gain depends only on other subsets.

For the rest of this chapter, we work by assuming that conjecture 3.1 is true. Having said that, let us have a collateral gain $r_0$ which depends on one or more subsets. We introduce the notation $r_1 \to r_0$ where $r_1$ is a variable representing a subset on which $r_0$ depends and we read "$r_0$ depends on $r_1$"; we call $r_1 \to r_0$ a (finite) chain of length 1. If $r_0$ depends on $k$ subsets, we can form $k$ different chains of length 1, one for each value of $r_1$. Now, subset $r_1$ can only be a determined subset (according to 3.1), so it is a checked subset or a collateral gain itself (according to corollary 2.6). If $r_1$ is a collateral gain, it depends on other subsets and therefore, the chain is extended backwards and takes the form $r_2 \to r_1 \to r_0$, which is a chain of length 2 and where $r_2$ is a subset on which $r_1$ depends. On the other hand, if $r_1$ is a checked subset, the chain cannot be extended backwards and comes to an end, since $r_1$ is determined independently of other subsets. As long as a collateral gain emerges, the chain is extended backwards, so let us give the general definition.

**Definition 3.3** With notation as above, we define a (finite) chain of length $m$ as

$$r_m \to r_{m-1} \to \cdots \to r_1 \to r_0 := (r_m \to r_{m-1}) \text{ and } (r_{m-1} \to r_{m-2}) \text{ and } \ldots \text{ and } (r_1 \to r_0)$$

where, for $k = 0, 1, 2, \ldots, m-1$, subset $r_k$ is a collateral gain with $r_k \not\equiv r_{k+1}$, while $r_m$ is a checked subset.

Let us note two things. Firstly, the requirement $r_k \not\equiv r_{k+1}$ is included so that the definition is in compliance with remark 3.2. For a collateral gain there are one or more chains attached to it, with each chain showing a path of dependencies followed by the algorithm in order to determine it; for each chain, the $r_k \not\equiv r_{k+1}$ requirement must be satisfied. Secondly, the sequence in which the dependencies are written in a chain indicates the priority order between them, in the sense of a dependence existing prior to another dependence. For example, for $r_{m-1} \to r_{m-2}$ ($r_{m-1}$ being a determined subset) to have meaning, we firstly need $r_m \to r_{m-1}$ for $r_{m-1}$ to be determined.

*Remark 3.4* Let us present the two following extensions of definition 3.3:

(1) A checked subset $r$, as we know, does not depend on other subsets in order to be determined. We can say that it depends only on itself and by using the notation of the definition, we can write $r \to r$. Here, $r$ on the left side is an undetermined subset, meaning that $r$ on the right side depends on an undetermined subset. This, of course, is permitted, since $r$ is a checked subset and not a collateral gain which would only then contradict 3.1.
(2) In the common case, an algorithm views the subsets of an input-set as $2^n - 1$ individual subsets. This is obvious in examples 2.2 and 2.3. But in example 2.7, we saw that the collateral gains are viewed in a different way; not individually but in sets. For $i = 2, 3, \ldots, n$, the algorithm determines all subsets of $x_1, \ldots, x_i$ except $\{x_i\}$ (as collateral gains) in one set, the $C_i \backslash \{x_i\}$ set (using the notation of the example). That is why, for this case in 2.7 and for any other possible case, we consider an extended version of definition 3.3, in which

[8]

$r_k$ (for $k = 0, 1, 2, ..., m$ ) can be a set of subsets. Back in example 2.7, we now have, for $i = 2, 3, ..., n$, the (finite) chain of length $i - 1$

$$\{x_1\} \to C_2 \backslash \{x_2\} \to C_3 \backslash \{x_3\} \to \cdots \to C_{i-1} \backslash \{x_{i-1}\} \to C_i \backslash \{x_i\}.$$

We should remember that for each $i$, the set $C_i \backslash \{x_i\}$ of collateral gains depends on the subsets of $x_1, ..., x_{i-1}$, which are not only the set $C_{i-1} \backslash \{x_{i-1}\}$ of collateral gains but also the checked subset $\{x_{i-1}\}$. So for a complete representation of all the paths of dependencies, in addition to the above chains, for $i = 3, ..., n$ we have

$$\{x_{i-1}\} \to C_i \backslash \{x_i\}.$$

Before we present the following proposition, we need to explain in advance the meaning of a certain phrase: by saying an algorithm has a finite or infinite (of no finite length) chain we will mean that such a chain (path of dependencies) appears in the way the algorithm operates.

**Proposition 3.5** *(assuming conjecture 3.1) An algorithm which determines all subsets of an input-set cannot have an infinite chain.*

*Proof* Let us assume that an algorithm can have an infinite chain and let that chain be $\cdots \to r_m \to r_{m-1} \to \cdots \to r_1 \to r_0$ where $r_k$ ($k = 0, 1, 2, ...$) is a collateral gain and $r_k \not\equiv r_{k+1}$. We consider the initial part $\cdots \to r_w \to r_{w-1} \to \cdots \to r_1 \to r_0$ of the chain, where $w = 2^n - 1$ ($n$ being the number of integers of the input-set $S$ ). But as we know, the number of different subsets[3] in our problem is finite and equal to $2^n - 1$. According to the pigeonhole principle, there must be a subset that appears at least twice in this initial part of the chain. More rigorously, there must be $i, j \in \{0, 1, 2, ..., w\}$ where $|i - j| > 1$ such that $r_i \equiv r_j$.

Let $e$ be the subset that appears at least twice and let $\cdots \to e \to \cdots \to d \to e \to \cdots$ be the context in which it appears, where $d$ is another subset. Then, $e$ can be determined only after $d$ is determined, according to $d \to e$ and 3.1. As a result, the $e$ in $e \to \cdots \to d$ cannot be determined, since it exists prior to $d$ being determined. But according to 3.1, the same $e$ must be determined, since there is a collateral gain that depends on it. This is a contradiction.

(We see that in $\cdots \to e \to \cdots \to d \to e \to \cdots$ both $e$ and $d$ depend on each other in order to be determined. This is similar to what happens in an equation of the form $x + y = a$, where $x, y \in \mathbb{Z}$ are the unknowns and $a \in \mathbb{Z}$ is a parameter. In such an equation, both $x$ and $y$ depend on each other in order to be determined as specific numbers and so, we cannot solve it for one unknown without knowing the value of the other. What we can do is to give arbitrarily a value to $x$ and then we can easily determine the value of $y$ which depends on $x$; but giving arbitrarily a value to $x$ means that $x$ is determined independently of $y$. Similarly, determining one subset independently of the other would be contradictory to $\cdots \to e \to \cdots \to d \to e \to \cdots$. Also, we can consider simultaneously two values, one for $x$ and one for $y$, and examine if the pair $(x, y)$ is a solution. But such an approach means that $x$ and $y$ do not depend on each other, since the values are really considered independently of

---

[3] For the proof, we use the common case of how an algorithm views the subsets of an input-set, that is as $2^n - 1$ individual subsets (see 3.4, (2)). This way of viewing is the simplest one and, exactly for this reason, the one that can (theoretically) result in the largest possible number of different collateral gains (all $2^n - 1$ subsets being collateral gains), larger than any other way of viewing. For example, in 2.7, the way the algorithm views the collateral gains is in sets and we saw that we can only have $n - 1$ different sets of collateral gains ($C_i \backslash \{x_i\}$ for $i = 2, 3, ..., n$). Therefore, the critical point $w = 2^n - 1$, which we later use in the proof in order to apply the pigeonhole principle, can also be used as a critical point to apply the pigeonhole principle for all ways of viewing. We could use even smaller critical points (like $w' = n - 1$ for 2.7-like ways of viewing), but $w$ is appropriate for a universal proof for all ways of viewing.

each other. Similarly, any approach to simultaneously determine subsets $e$ and $d$ would mean that they no longer depend on each other; they may depend on other subsets (still being collateral gains), but not on each other in such a case. And that would again be contradictory to $\cdots \to e \to \cdots \to d \to e \to \cdots$ ) □

Note that we proved proposition 3.5 without making any reference to the steps of an algorithm. We proved it based only on the nature of the dependencies.

So, we showed that an algorithm cannot have an infinite chain. Therefore, an algorithm can have only finite chain(s). Now, let $r_m \to r_{m-1} \to \cdots \to r_1 \to r_0$ be a (finite) chain of length $m$, where checked subset $r_m$ is considered to be the starting point of the chain. We make the following observation.

*Remark 3.6* For a collateral gain to be determined, the starting point of each of its chains has to be a checked subset.

As an application, we see that in example 2.3, the starting point is always a "current couple of elements" which is a checked subset and in example 2.7, the starting point is always a $\{x_i\}$ ( $i = 1, 2, \ldots, n-1$ ) which is a checked subset.

So, as far as collateral gains are concerned, everything has to start with a checked subset. This shows how important checked subsets are for the algorithmic process.

Let $P$ be a polynomial-time algorithm. If $P$ was able to check at most one subset in a single step ($P_1$-property), the total number of checked subsets would be polynomially bounded. But the number of subsets to be determined increases exponentially ($2^n - 1$). To compensate for the increasing difference between the two numbers, $P$ would have to rely more and more heavily on collateral gains. But collateral gains themselves, in order to be determined, need checked subsets (according to remark 3.6). Given that there is a limit to how much information can be derived from a checked subset (even from checked subsets combined), the polynomially bounded number of checked subsets would eventually (as $n$ increases) not be enough for producing the entire, exponentially increasing number of collateral gains. Therefore, we believe that an algorithm cannot reach polynomial "speeds" by checking at most one subset in a single step. It needs to be able to check more than one subsets in a step. And that was the reasoning behind conjecture 2.8.

**4 The normal and the problematic nature of the algorithms**

In this section, we describe the theoretical mathematical environment we talked about in the introduction. But before we do that, we need to provide the general setup, so that we can configure some technical details that arise when dealing with algorithms in general and $P_1/P_2$-algorithms in particular.

**Setup 4.1** This general setup consists of the following three notes:

(1) The number of steps of an algorithm, for all input-sets of the same size (cardinality), is considered to be the number of steps dictated by the time the worst-case algorithmic scenario requires for input-sets of this size. In that way, we work with a common base – in the number of the steps – which refers to all input-sets of the same size.

In example 2.2, it is obvious that the number of steps of the brute-force algorithm is always $2^n - 1$; even when a subset-solution is found in a step which is earlier than the $(2^n - 1)^{th}$ (last step), the algorithm performs its entire operation of determining all subsets in $2^n - 1$ steps.

In other algorithms, like in example 2.3, the number of steps (that the process of determining all subsets requires) varies depending on the input-set (we always refer to input-sets of the same size). So, for an algorithmic process that requires fewer steps than the worst-case scenario, we add empty steps to it until the total number reaches that of the worst-case scenario. By doing so, we do not affect the algorithmic process in any way, since the steps are added after its completion and the algorithm performs no operations at all in them (empty steps), and we achieve the common base/number we talked about. This is similar to adding zeros in the decimal places of a number, for example the number 2,75 can be equivalently written as 2,7500.

(2) Different input-sets of the same size may result in different distributions of checked subsets into the steps of the algorithm. By "distribution of checked subsets into the steps of the algorithm" we refer to an ordered $k$-tuple $d = (d_1, d_2, ..., d_k)$, where $k$ is the number of the steps of the algorithm and $d_j$ ($j = 1, 2, ..., k$) is the number of checked subsets in the $j^{th}$ step (by that, of course, we mean the number of subsets the algorithm checks in the $j^{th}$ step). Two input-sets result in the same distribution if and only if they result in equal $k$-tuples. In $P_1$-algorithms, $d_j$ can only be 0 or 1 (by definition), whereas in $P_2$-algorithms, $d_j$ could be a number greater than 1 (again by definition).

In example 2.2, if we consider that a subset is checked in each step, then the distribution is always the same and in particular, one checked subset in each step. But in example 2.3, things are different. The empty steps that we add to the algorithmic process translate into a sequence of zeros in the ending of the $k$-tuple (since the algorithm performs no operations in these steps, thus no/zero subsets are checked). And different input-sets may result in different numbers of empty steps needed to be added, therefore different sequences (different in length) of zeros in the endings of their $k$-tuples, therefore different distributions.

In $P_2$-algorithms, where more than one subsets are expected to be checked in a main step, things could be even more complicated.

However, for both $P_1$ and $P_2$-algorithms and for input-sets of the same size, all possible distributions of checked subsets into the steps of an algorithm can only be finite in number, since both the steps and the subsets are finite. On the other hand, the number of different input-sets (of the same size) is infinite. So, by the pigeonhole principle, there is at least one distribution which is the same for an infinite number of input-sets of the same size.

Therefore, we consider the following. Let $V$ be a $P_1$ or $P_2$-algorithm and $V_{input}(n, d)$ the class which contains an infinite number of input-sets of size $n$ that result in the same distribution $d$ of checked subsets into the steps of $V$. We call $V_{input}(n, d)$ input-class of $V$.

So, for each input-set that belongs to $V_{input}(n, d)$, algorithm $V$ has the checked subsets distributed the same way (denoted by $d$) into its steps. By selecting input-sets from $V_{input}(n, d)$, we ensure that, as far as checked subsets are concerned, a common distribution is used when dealing with the algorithm.

(3) In $P_1$-algorithms, there is only one checked subset in a main step. So, if a subset-solution is to be determined independently of other subsets in a specific main step, there is only one place where it can be determined and that place is the (only) subset that is checked. In $P_2$-algorithms, however, if a subset-solution was to be determined independently in a main step where $\ell$ subsets were checked ($\ell > 1$), then we would obviously have $\ell$ possible places (one for each checked subset).

Now, before we continue, let us remember what happens in a classical random experiment, that of rolling two dice. The sample space of this random experiment is $\Omega = \{(1,1), (1,2), ..., (4,6), ..., (6,4), (6,5), (6,6)\}$. The fact that simple events like $(4,6)$ and $(6,4)$ are both included in $\Omega$ means that we consider them as two different simple events and not the same one. By naming/labeling one dice as the "first" and the other as the "second", simple event $(4,6)$ means that the indication of the "first" dice is 4 and the indication of the "second" is 6, whereas $(6,4)$ means the opposite. Of course, instead of "first" and "second", we could use other names like "green" dice and "red" dice, always in order to view each dice individually. Whatever the names/labels, that individuality ensures that the sample space is completed correctly.

That is exactly why we must name/label the $\ell$ possible places (existing in the $i^{th}$ step of a $P_2$-algorithm) by enumerating them. So that we can view them as individual and be able to say that the event

{ $a\ subset - solution\ is\ determined\ independently\ of\ other\ subsets\ in\ the\ i^{th}\ step$ }

is satisfied by simple events that have a subset-solution being determined in at least one of the $\ell$ places (for example, a subset-solution being determined in the $1^{st}$ place, or in the $3^{rd}$ place, or two subsets-solutions being determined, one in the $2^{nd}$ and one in the $\ell^{th}$ place etc.).

That was the general setup we needed to provide so that we can now describe the mathematical environment in which the normal nature of $P_1$-algorithms and the problematic nature of $P_2$-algorithms appear.

The basic tool in this environment is a new kind of stochastic process. As we know, a stochastic process describes a system/phenomenon/random experiment and its evolution over an axis ($\mathbb{R}$-axis or $\mathbb{N}$-axis) which usually represents time, space or something else. Here, we introduce an axis of a different kind, an axis which represents the probability of an event. More specifically, we focus on a step (of the algorithm we deal with) and the axis represents the probability of a subset-solution being determined independently (of other subsets) in this step. As this probability increases (just like time would advance in a typical stochastic process), we study the evolution of the process over the probability. Additionally, the way this probability is generated results in the axis being neither an $\mathbb{N}$-axis nor an $\mathbb{R}$-axis, but a $\mathbb{Q}$-axis.

The reasons discussed above justify the term "new" in what we called "a new kind of stochastic process". Now, let us start by presenting the system/random experiment that the stochastic process is going to describe. We should note that the random experiment and the entire mathematical environment that follow are theoretical, in the sense that no physical constraints are considered.

Let us have a $P_1$-algorithm $V$, an input-class $V_{input}\ (n,d)$ and let us consider a main step (as indicated by distribution $d$) of $V$, the $i^{th}$ step. Now, we select two input-sets from $V_{input}\ (n,d)$ with the following properties: an input-set $S$ which results in a subset-solution being determined as checked subset (independently of other subsets) in the $i^{th}$ step and an input-set $T$ which results in no subset-solution being determined as checked subset in the $i^{th}$ step (either all subsets-solutions that are determined as checked subsets are determined in steps other than the $i^{th}$ or no subset-solution is determined as checked subset at all).

*Remark 4.2* The selection of the two input-sets $S$ and $T$ is possible for the following reasons:

(1) As we know, an algorithm is a finite procedure. Therefore, given an input-set, we can always run the algorithm and find out the step in which a subset-solution is determined for this input-set.
(2) The fact that the subset-solution is asked to be determined independently of other subsets ensures that input-sets like $S$ and $T$ exist. Let us explain what we mean by that. There are algorithms where a collateral gain is denied the possibility of being a subset-solution, whatever the input-set, due to the fact and the way that it depends on other subsets. Such a case is in example 2.3 where a subset-combination/collateral gain emerges only to be excluded from a potential subset-solution, for all input-sets. For a checked subset, on the other hand, we can always expect it either to be a subset-solution or not, since it does not depend on anything. Therefore, the – fixed by distribution $d$ – place of the checked subset in the $i^{th}$ step (which hosts a subset in each algorithmic run) must be able to host subsets that are subsets-solutions as well as subsets that are not, depending on the input-set; if one of these options was denied, it would not agree with the independent way in which the subset was determined. So, there must be an input-set $S$ which results in a subset-solution being determined as checked subset in the $i^{th}$ step, as well as an input-set $T$.

Now, let us consider a set $B = \{S, T\}$. If we randomly select an input-set from $B$ to run algorithm $V$ with, the probability of the event

$A = \{$ a subset $-$ solution is determined independently of other subsets in the $i^{th}$ step $\}$

is $P(A) = \frac{1}{2}$, since there are two input-sets in $B$ and only one of them, $S$, satisfies event $A$.

Similarly, if we consider a multiset $B' = \{S, S, T\}$ and randomly select an input-set from $B'$, it is $P(A) = \frac{2}{3}$.

Generally, we define a multiset $Ba_{/b}$, for $a, b \in \mathbb{N}$, $b \neq 0$ and $a \leq b$, as

$$Ba_{/b} = \{S, S, S, \ldots, S, T, T, T, \ldots, T\},$$

where the number of instances for $S$ is $a$ and the number of instances for $T$ is $b - a$. We can easily see that the cardinality (the total number of instances) of $Ba_{/b}$ is $b$. A multiset $Ba_{/b}$ will be called box $Ba_{/b}$.

Therefore, in the earlier examples that we offered, set $B$ and multiset $B'$ are respectively box $B1_{/2}$ and box $B2_{/3}$. As far as the name "box" is concerned, we chose it for two reasons. Firstly, a box $Ba_{/b}$ is a multiset of sets (input-sets), as we defined it to be. So, one reason is to avoid phrases like "a multiset of sets" and even "a set of multisets of sets" (which would be probably used). Secondly and more importantly, in the context of the random experiment of randomly selecting an input-set from a box $Ba_{/b}$, we can visualize $Ba_{/b}$ as being a real box containing $b$ (equally likely to be selected) input-sets.

*Remark 4.3* Let us make the following three important notes:

(1) We saw that for box $B1_{/2}$ we have $P(A) = \frac{1}{2}$ and for box $B2_{/3}$ we have $P(A) = \frac{2}{3}$. It is clear that in general, if we randomly select an input-set from box $Ba_{/b}$ to run algorithm $V$ with, it is $P(A) = \frac{a}{b}$.

(2) Since $a, b \in \mathbb{N}$, $b \neq 0$ and $a \leq b$ (by definition of $Ba_{/b}$), the fraction $\frac{a}{b}$ will always be equal to a rational number $q$ where $0 \leq q \leq 1$. More specifically, for each $q \in \mathbb{Q} \cap [0,1]$ there is an infinite family of boxes $Ba_{/b}$ such that $\frac{a}{b} = q = P(A)$ (for example, for $q = 0.5$ we have $B1_{/2}, B2_{/4}, B3_{/6}, B4_{/8}, \ldots$).

(3) By appropriately changing the numbers of instances for $S$ and $T$ (essentially, numbers $a$ and $b$), we can repeatedly perform our random experiment (of randomly selecting an input-set from a box) using a sequence of boxes that result in an increasing probability of event $A$. In that way, an axis of probability is generated over which our random experiment evolves, just like a random experiment would more commonly evolve over the axis of time.

We are ready to present the stochastic process now. In order to achieve a more elegant presentation, we work as follows. Instead of saying "input-set $S$ results in a subset-solution being determined as checked subset in the $i^{th}$ step", we will be saying "input-set $S$ results in a subset-solution being determined as checked subset in the $1^{st}$ (and only) place of the $i^{th}$ step". And instead of "input-set $T$ results in no subset-solution being determined as checked subset in the $i^{th}$ step", we will be saying "input-set $T$ results in a subset-solution being determined as checked subset in the $0^{th}$ place of the $i^{th}$ step".

For $q = \frac{a}{b} \in \mathbb{Q} \cap [0,1]$, let $X_q$ be a random variable which takes the following values:

[13]

$$X_q = \begin{cases} 0 & \begin{array}{l}\text{if the input}-\text{set, that is randomly selected from box } Ba_{/b} \text{ with } P(A) = q, \\ \text{results in a subset}-\text{solution being determined as checked subset} \\ \text{in the } 0^{th} \text{ place of the } i^{th} \text{ step of } V. \end{array} \\ \\ 1 & \begin{array}{l}\text{if the input}-\text{set, that is randomly selected from box } Ba_{/b} \text{ with } P(A) = q, \\ \text{results in a subset}-\text{solution being determined as checked subset} \\ \text{in the } 1^{st} \text{ place of the } i^{th} \text{ step of } V. \end{array} \end{cases}$$

The collection $\{X_q\}$ of random variables is a stochastic process with index set $\mathbb{Q} \cap [0,1]$ and state space $\{0,1\}$. As $q \to 1$ (taking only rational values, as dictated by the index set) the probability of event $A$ increases, forming the axis of probability. Each rational value of $q$ is produced by randomly selecting from a box $Ba_{/b}$ which can be any box as long as $\frac{a}{b} = q$ (remark 4.3, (2), about the "infinite family of boxes").

Now, we define the following probabilities:

$$p_1(q) = prob\{X_q = 1\} \quad \text{and} \quad p_0(q) = prob\{X_q = 0\}, \quad \text{for } q \in \mathbb{Q} \cap [0,1] .$$

These probabilities are a simple but very important tool when studying the stochastic process. It is through this tool that we are going to reveal the problematic nature of $P_2$-algorithms later in the paper, in contrast to the normal nature of $P_1$-algorithms.

We can easily see that

$$p_1(q) = q \quad \text{and} \quad p_0(q) = 1 - q , \quad \text{for } q \in \mathbb{Q} \cap [0,1] .$$

Indeed, we know by definition that $X_q = 1$ if the input-set, that is randomly selected from box $Ba_{/b}$ with $P(A) = q$, results in a subset-solution being determined as checked subset in the $1^{st}$ place of the $i^{th}$ step. Equivalently, $X_q = 1$ if the input-set, that is randomly selected from box $Ba_{/b}$ with $P(A) = q$, is $S$. Therefore, $p_1(q) = \frac{a}{b} = q$. Similarly, but for input-set $T$ instead of $S$, we have $p_0(q) = 1 - q$.

And here comes the main result with which we want to justify the normal nature of $P_1$-algorithms.

*Remark 4.4* The probability $p_1(q): \mathbb{Q} \cap [0,1] \to [0,1]$, $p_1(q) = q$, is continuous.

This result may seem trivial, but its strength will be fully appreciated after the comparison to $P_2$-algorithms.

For the probability of event $A$ to be produced and calculated in the context of our random experiment, we need to know the number of input-sets that satisfy event $A$, which is number $a$ (the number of instances for $S$ in a box $Ba_{/b}$). In $P_1$-algorithms, number $a$ can be filled – and therefore, event $A$ can be satisfied – only by one type of input-sets, the type that results in a subset-solution being determined as checked subset in the only place available of the $i^{th}$ step. The "$S$-like" type if you want, using input-set $S$ as a representative.

In $P_2$-algorithms, on the other hand, things are different. As we talked about in 4.1, (3), we can find a main step where there are $\ell$ available places ($\ell > 1$) for a subset-solution to be determined as checked subset. In the analysis that follows, for simplicity, we work for $\ell = 2$.

Let us have a $P_2$-algorithm $V'$, an input-class $V'_{input}(n, d')$ and let us consider, as indicated by distribution $d'$, a main step (of $V'$) in which $\ell = 2$ subsets are checked, the $i^{th}$ step (not to be confused with the $i^{th}$ step of $V$, we are just using the same letter $i$ for convenience). Now, we select three input-sets from $V'_{input}(n, d')$ with the following properties: an input-set $S'$ which results in a subset-solution being determined as checked subset in the $1^{st}$ place of the $i^{th}$ step but no subset-solution being determined in the $2^{nd}$ place, an input-set $R$ which results in a subset-solution being determined as checked subset in the $2^{nd}$ place of the $i^{th}$ step but no subset-solution being determined in the $1^{st}$ place and an input-set $T'$ which results in a subset-solution being determined as checked subset in the $0^{th}$ place of the $i^{th}$ step (again, as in the analysis of $V$, the $0^{th}$ place means that no subset-solution is determined as checked subset in the $i^{th}$ step).

Let us note that even if we did not mention it (because we are not going to need it), there is also another type of input-sets that satisfies event $A$ apart from $S'$ and $R$, the type that results in two subsets-solutions being determined as checked subsets in the $i^{th}$ step, one in the $1^{st}$ place and one in the $2^{nd}$ place.

*Remark 4.5* The selection of the three input-sets $S', R$ and $T'$ is possible for the following reasons:

(1) For the same reasons that are described in remark 4.2, (1), (2).
(2) If having a subset-solution determined in one place of the $i^{th}$ step meant that a subset-solution is also determined in the other place, for all input-sets, then again this would not agree with the independent way in which the subsets were determined. Therefore, there must be an input-set $S'$ which results in a subset-solution being determined as checked subset in the $1^{st}$ place of the $i^{th}$ step but no subset-solution being determined in the $2^{nd}$ place, as well as an input-set $R$.

Now, let us give a more general definition of a box. We define a multiset $B'_{a_1, a_2/b}$, for $a_1, a_2, b \in \mathbb{N}$, $b \neq 0$, $a_1 + a_2 = a$ and $a \leq b$, as

$$B'_{a_1, a_2/b} = \{S', S', S', \ldots, S', R, R, R, \ldots, R, T', T', T', \ldots, T'\},$$

where the number of instances for $S'$ is $a_1$, the number of instances for $R$ is $a_2$ and the number of instances for $T'$ is $b - a$. We can easily see that the cardinality (the total number of instances) of $B'_{a_1, a_2/b}$ is $b$. A multiset $B'_{a_1, a_2/b}$ will be called box $B'_{a_1, a_2/b}$. The definition is more general in the sense that we could also use the notation to describe a box $B_{a/b}$ (found in the analysis of $V$), writing it as $B_{a, 0/b}$.

We consider the random experiment of randomly selecting an input-set from a box $B'_{a_1, a_2/b}$ and we also consider the same event $A$ that was used in the analysis of $V$. Again, we see that if we randomly select an input-set from box $B'_{a_1, a_2/b}$ to run algorithm $V'$ with, it is $P(A) = \frac{a}{b}$, since $S'$ and $R$ both satisfy event $A$ and $a_1 + a_2 = a$. In general, the entire remark 4.3 is also true here, with the obvious necessary adjustments so that we refer to $V'$ and boxes $B'_{a_1, a_2/b}$.

Now, for $q = \frac{a}{b} \in \mathbb{Q} \cap [0,1]$, let $X'_q$ be a random variable which takes the following values:

$$X'_q = \begin{cases} 0 & \text{if the input}-\text{set, that is randomly selected from box } B'_{a_1,a_2/b} \text{ with } P(A) = q, \\ & \text{results in a subset}-\text{solution being determined as checked subset} \\ & \text{in the } 0^{th} \text{ place of the } i^{th} \text{ step of } V'. \\ \\ 1 & \text{if the input}-\text{set, that is randomly selected from box } B'_{a_1,a_2/b} \text{ with } P(A) = q, \\ & \text{results in a subset}-\text{solution being determined as checked subset} \\ & \text{in the } 1^{st} \text{ place of the } i^{th} \text{ step of } V'. \\ \\ 2 & \text{if the input}-\text{set, that is randomly selected from box } B'_{a_1,a_2/b} \text{ with } P(A) = q, \\ & \text{results in a subset}-\text{solution being determined as checked subset} \\ & \text{in the } 2^{nd} \text{ place of the } i^{th} \text{ step of } V'. \end{cases}$$

The collection $\{X'_q\}$ of random variables is a stochastic process with index set $\mathbb{Q} \cap [0,1]$ and state space $\{0,1,2\}$. As $q \to 1$, the probability of event $A$ increases, forming the axis of probability.

We define the following probabilities for $q \in \mathbb{Q} \cap [0,1]$ :

$$\pi_2(q) = prob\{X'_q = 2\}, \quad \pi_1(q) = prob\{X'_q = 1\} \quad \text{and} \quad \pi_0(q) = prob\{X'_q = 0\}.$$

We used the greek letter $\pi$ to define the probabilities – from the initial letter of the greek word for probability – in order to distinguish them from the probabilities in the analysis of $V$.

We can easily see that for $q \in \mathbb{Q} \cap [0,1]$, it is

$$\pi_2(q) = \frac{a_2}{b}, \quad \pi_1(q) = \frac{a_1}{b} \quad \text{and} \quad \pi_0(q) = \frac{b-a}{b} = 1 - q.$$

As we know, each value of $q$ is produced by randomly selecting an input-set from a box $B'_{a_1,a_2/b}$ which can be any box as long as $\frac{a_1+a_2}{b} = \frac{a}{b} = q$. For example, for $q = 0.5 = \frac{1}{2} = \frac{2}{4} = \frac{3}{6} = \cdots$, we can randomly select an input-set from whichever box of the following: $B'_{1,0/2}, B'_{0,1/2}, B'_{1,1/4}, B'_{2,0/4}, B'_{0,2/4}, \cdots$.

Now, we consider an example of a family $F$ of boxes which produce the entire index set. The boxes that $F$ contains are given by the following rule:

$$\begin{cases} B'_{a,0/b}, & \text{for } q \in \mathbb{Q} \cap [0,1) \\ & \text{and} \\ B'_{0,b/b}, & \text{for } q = 1 \end{cases}$$

With this rule, the boxes that produce the values of $q \in \mathbb{Q} \cap [0,1)$ do not contain instances of input-set $R$, whereas the box for $q = 1$ contains only instances of $R$ (it can contain one or more instances of $R$, since either case produces the value $P(A) = q = 1$).

By using the boxes of $F$ for producing all values of $q \in \mathbb{Q} \cap [0,1]$ (the entire index set), the probability $\pi_1(q)$ takes the following values:

[16]

$$\pi_1(q) = \frac{a_1}{b} = \begin{cases} \frac{a}{b} = q & if \ q \in \mathbb{Q} \cap [0,1) \\ \\ 0 & if \ q = 1 \end{cases}$$

We see that the probability $\pi_1(q): \mathbb{Q} \cap [0,1] \to [0,1]$, when using the boxes of $F$, is discontinuous at the point $1$. And $F$ is just one example. By exploiting the fact that both $S'$ and $R$ satisfy event $A$, as we did in $F$, we can consider a family of boxes (which produce the entire index set) that results in the $\pi_1(q)$ being discontinuous at another point, or a family that results in the $\pi_1(q)$ being discontinuous at two or more points; if we are creative enough with the rule that gives the boxes, we can even consider a family that results in the $\pi_1(q)$ being discontinuous at an infinite number of points. (Similarly for probability $\pi_2(q)$)

These discontinuities are what we mean by saying that the nature of $P_2$-algorithms is problematic. And let us further explain this in the paragraphs that follow.

From a technical point of view, the fact that discontinuities are allowed to appear (depending on the family of boxes that is used) makes the stochastic process mathematically "exotic". In the analysis of stochastic processes, when similar probabilities appear (transition probabilities in homogeneous Markov processes, in Poisson processes, in Queueing systems etc.), we are used to working with continuous –and not discontinuous– functions that represent these probabilities. Such functions, being continuous, provide mathematical convenience and richness in the analysis. But in the stochastic process $\{X'_q\}$, the probability $\pi_1(q)$ can be discontinuous due to property $P_2$ allowing for more than one subsets to be checked in the same step.

From a mathematically philosophical point of view, it is the natural meaning of $\pi_1(q)$ as a probability that is not compatible with being discontinuous. This is clear not only in the case where $\pi_1(q)$ is discontinuous at an infinite number of points, but even in the relatively "simple" case where it is discontinuous at a single point, like with family $F$. When using the boxes of $F$ for producing all values of $q$, then as $q \to 1$, the probability $\pi_1(q) = q$ becomes arbitrarily large, in other words as close to $1$ as we want; but for $q = 1$, the probability $\pi_1(q)$ is zero. Given that $q \in \mathbb{Q} \cap [0,1]$ and that there is no notion of immediately preceding/following number in rational numbers (densely ordered), the jump from "as close to $1$ as we want" to zero becomes even more difficult to be meaningfully interpreted.

Finally but most importantly, it is through the comparison to $P_1$-algorithms that the problematic nature of $P_2$-algorithms becomes even more significant. In $P_1$-algorithms, we saw that the probability $p_1(q)$ is continuous (remark 4.4). Exactly due to property $P_1$, there is no way for a stochastic process $\{X_q\}$ to generate a discontinuity for $p_1(q)$; in other words, this property of continuity is the only scenario. And we believe that it is very important that this normal behavior of continuity is found when dealing with $P_1$-algorithms, where all known exponential-time algorithms are contained (Fig.1). On the other hand, it is in $P_2$-algorithms, where no algorithm is known to exist, that we have the problematic nature of discontinuities.

(For all the previous results, let us note that we could also use another random variable for the stochastic processes, instead of $X_q$ and $X'_q$. That random variable would be $Y_q$, expressing the number of subsets-solutions being determined as checked subsets in the $i^{th}$ step, as a result of the input-set that has been randomly selected from a box with $P(A) = q$; such a random variable would take the values $0,1,2$. We would consider the same event $A$, for the axis of probability, and appropriate input-sets, for the boxes to contain, so that the random variable would take the above values)

## 5 Final remarks

The word "problem" comes from the greek word "πρόβλημα" which means any issue whose solution is characterized by difficulties. Here, as far as the problematic nature of $P_2$-algorithms is concerned, we described the difficulties through a technical, a mathematically philosophical and a comparison-driven (comparison to $P_1$-algorithms) point of view. The reason why these difficulties appear is either the fact that the class of $P_2$-algorithms cannot exist or it is too mathematically "exotic".

The use of the stochastic processes as a tool was to demonstrate the problematic nature of $P_2$-algorithms as a result of their fundamental difference to $P_1$-algorithms in the number of checked subsets in a step. In order to strictly show whether $P_2$-algorithms exist, we should question whether checking more than one subsets in a step is possible. We believe that it is impossible, being an idea that is not compatible with the mathematical and the physical reality. Given the conjecture we presented (2.8), the non-existence of $P_2$-algorithms would mean that the polynomial-time algorithms, that determine all subsets of an input-set, do not exist.

Finally, let us highlight the importance of the two classifications around which the entire paper revolves. The one is the $P_1$/ $P_2$-algorithms, based on the number of checked subsets in the same step. This classification provides the context in which the comparison takes place, as well as the (conjectured) link with the exponential/polynomial-time algorithms solving the subset-sum problem. The other classification is the checked subset/collateral gain, based on the independent or not way in which a subset is determined by the algorithm. The element of independence is very important, because it makes the use of the stochastic processes possible by unlocking the selection of the input-sets contained in the boxes, as we discussed about in remarks 4.2, (2) and 4.5.